# Req-Rec: Enhancing Requirements Elicitation for Increasing Stakeholder's Satisfaction Using a Collaborative Filtering Based Recommender System

**Ali Fallahi RahmatAbadi [a], Amineh Amini [a*], Azam Bastanfard [a], Hadi Saboohi [a]**

[a] Department of Computer Engineering, Karaj Branch, Islamic Azad University, Karaj, Iran, Corresponding author: aamini@kiau.ac.ir

**Abstract**

The success or failure of a project is highly related to recognizing the right stakeholders and accurately finding and discovering their requirements. However, choosing the proper elicitation technique was always a considerable challenge for efficient requirement engineering. As a consequence of the swift improvement of digital technologies since the past decade, recommender systems have become an efficient channel for making a deeply personalized interactive communication with stakeholders. In this research, a new method, called the Req-Rec (Requirements Recommender), is proposed. It is a hybrid recommender system based on the collaborative filtering approach and the repertory grid technique as the core component. The primary goal of Req-Rec is to increase stakeholder satisfaction by assisting them in the requirement elicitation phase. Based on the results, the method efficiently could overcome weaknesses of common requirement elicitation techniques, such as time limitation, location-based restrictions, and bias in requirements' elicitation process. Therefore, recommending related requirements assists stakeholders in becoming more aware of different aspects of the project.

## 1. Introduction

Requirements engineering (RE) is known as the most crucial section of software engineering for any successful project (Dabbagh et al., 2016; Palomares et al., 2018). Nowadays, the importance of deficient and inaccurate RE as the most significant reason for increasing the probability of projects' failure can no longer be ignored (Mishra et al., 2008; Pohl & Rupp, 2015). Many studies emphasized the importance of RE, especially in software projects (Alkhammash, 2020; Ramachandran, 2016). RE covers all the activities about: recognizing the stakeholders and understanding their requirements, analyzing and documenting the explorations, discussing the results, and finally implementing the proposed system (Melegati et al., 2019). These activities are considered as an engineering process because the main idea behind them is still providing useful and cost-effective solutions for realistic challenges (Nuseibeh & Easterbrook, 2000).

Requirements engineering includes four main core activities: elicitation, evaluation and agreement, negotiation and documentation, and the validation and release planning phase as the last activity (Felfernig et al., 2013). Elicitation is about understanding the domain of the system and discovering the stakeholder's needs and expectations. The main goal of the evaluation and agreement activities is to detect inconsistencies, conflicts, and also risks of the elicited requirements. Providing an understandable and unambiguous version of the requirements document (RD) needs some activities such as technical discussions and meetings with stakeholders, which are samples of the negotiation and documentation phase. Validation and release planning is the final step of the RE process, including reviewing the recruitment document to validate it for clarity, consistency, and completeness. The output of this phase is consolidated requirements that accurately represent the stakeholder's needs (Marcelino-Jesus et al., 2014). The requirement engineering process is iterative, and each step can cause some new requirements identified and added to the process. All the four mentioned core steps of requirement engineering and their relationship are shown in Fig. 1.

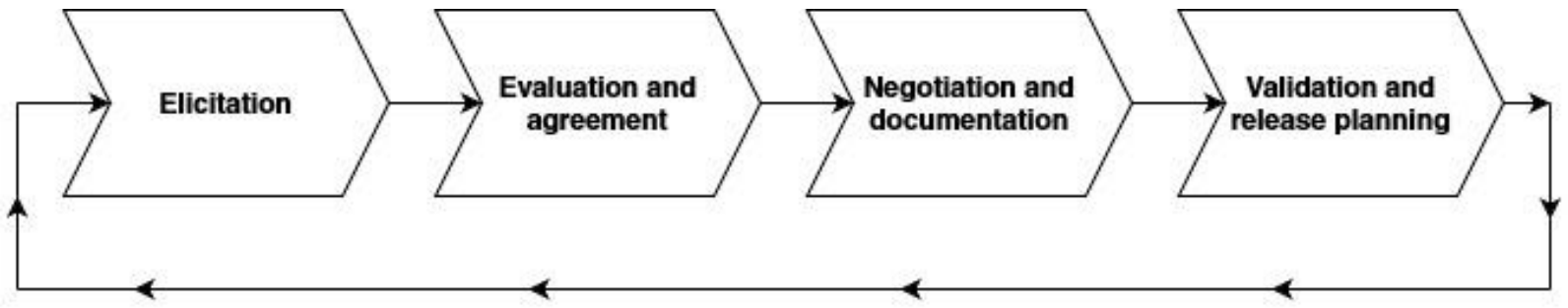

**Fig. 1. Requirement engineering's four core steps.**

All the Among all of the mentioned requirements engineering steps, elicitation is considered the core and fundamental activity (Pacheco et al., 2018; Pohl & Rupp, 2015). Based on the importance of this phase, many researchers chose the requirements elicitation and identification as the main subject of their studies and tried to propose an efficient method for improving the accuracy and precision of the elicited requirements (Silva et al., 2017; Wong et al., 2017). While there is a variety of requirement elicitation techniques, in general, all the methods can be categorized into four main categories: traditional, contextual, collaborative/group, and cognitive (Tiwari et al., 2012; Yousuf & Asger, 2015). Fig. 2 shows some of the most well-known requirement gathering techniques and approaches and their relations.

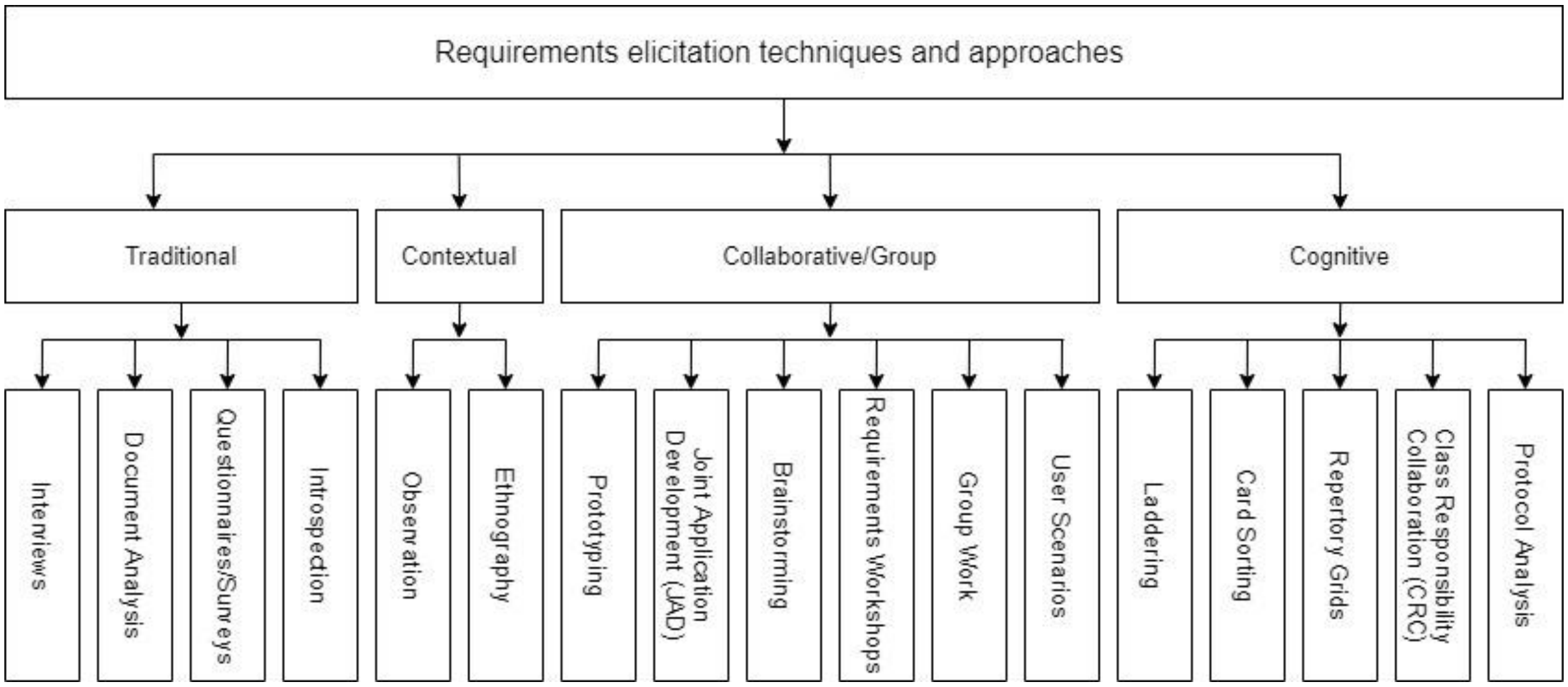

**Fig. 2. Requirement elicitation techniques and approaches.**

Although each of the introduced techniques in Fig. 2. has its pros and cons, none of them is utterly perfect for eliciting requirements in all conditions (Rafiq et al., 2017). Many studies indicated that there is not an absolutely ideal solution for identifying stakeholders' needs (A. Davis et al., 2006), and all the influential factors in a project such as available resources, time limitations, etc. have to be considered for choosing the eliciting method (Hickey & Davis, 2003; Nuseibeh & Easterbrook, 2000).

By considering the explained situation and investigating some of the most common elicitation techniques and their advantages and disadvantages, in the Req-Rec method first, the repertory grid technique was chosen as a way to discover what stakeholders want clearly. Repertory grid can be highly effective for evading common perceptual bias in requirements’ elicitation process. Moreover, compared to the techniques, the repertory grid provides precision results while also balancing the differences between stakeholders' and experts' points of view. Besides the mentioned values, the repertory grid concept is highly consistent with the user-item rating matrix, which is the core factor of recommender systems (Shaw, 1980). As the contribution of this study, a collaborative filtering-based recommender system is proposed, which uses an interactive repertory grid structure to provide an enhanced requirement elicitation technique. Recommender systems have long been a field of interest both in academic studies and business since the 1990s, when the first papers about using the collaborative filtering technique were published (Mazeh & Shmueli, 2020). However, the swift increment of the internet role since the past decade provided some new potentials in businesses that were not accessible in the past (Takbiri et al., 2019).

Recommending highly personalized content to answer the stakeholder's requirements and using social networks to improve the stakeholder's engagement provided brilliant opportunities for researchers and firms (Goyal & Goyal, 2021; Zhang et al., 2021). The reputation of recommender systems also has another reason, and that is because these systems have a tremendous potential to match with other fields of science and improve the stakeholder's satisfaction; by considering this feature, recommender systems have been widely applied in diverse areas such as e-commerce (Zhan & Xu, 2023), news (Karimi et al., 2018), music (Li et al., 2007), etc. The primary purpose of recommender systems is to assist stakeholders in finding what they require. In this way, a recommender acts as a decision support system, and by considering what is relevant to each stakeholder's needs, filter the available information, and provide personalized suggestions (Gómez et al., 2022). This feature has caused recommender systems to become useful in software engineering and its core activities, such as requirement engineering (Felfernig et al., 2013; Pakdeetrakulwong et al., 2014).

Lots of methods have been used in different studies to make recommender systems. However, the most popular of these methods can be categorized into content-based filtering, demographic, knowledge-based, collaborative filtering, and hybrid recommender systems (Roul & Arora, 2019). Among all of the mentioned techniques, collaborative filtering is the most common technique to build a recommender system (Alhijawi & Kilani, 2020; Kuo et al., 2021). Collaborative filtering recommender system stays efficient by considering user interests over time, so these kinds of systems cause enhancement in stakeholders' engagement, especially over a long period of interaction (de Campos et al., 2010). Moreover, the most considerable strength of collaborative filtering compared to the other mentioned techniques for building a recommender system is that understanding items' content is not mandatory in collaborative filtering.

Hence, the system has lots of flexibility to become compatible with different domains (Ajoudanian & Abadeh, 2019). Regarding the mentioned features, in this study, collaborative filtering (CF) is used to build the Req-Rec with the aim of increasing stakeholders' satisfaction by enhancing the elicitation phase activities. At the same time, it has no pre-defined limitation for eliciting requirements in various fields (Moon et al., 2013; Shaw, 1980).

The rest of this paper is organized as follows: Section 2 covers related works and studies, section 3 introduces our proposed method, section 4 explains our experimental results and evaluates the achievements; finally, section 5 concludes this study.

## 2. Related Work

Using the advantages of recommendation systems as surveyed and reviewed in (Al-walidi et al.; Ninaus, 2016; Williams & Yuan, 2019) has been a growing interest in different activities of the requirements engineering (RE). Mobasher and Cleland-Huang (Mobasher & Cleland-Huang, 2011) in their research about the role of recommender systems (RecSys) in RE, indicated the three certain types of issues, which are identifying the right stakeholders, determining and eliciting requirements, and finally supporting and helping users in other related activities such as requirements prioritization. Another valuable study about the RecSys functionality within the RE domain has been done with Maalej and Thurimella (Maalej & Thurimella, 2009); in the research, the authors described ideas about the benefits of recommenders in RE, such as improving saving times in background study, knowledge reuse, etc.

Likewise, Roher and Richardson (Roher & Richardson, 2013) worked on how using the RecSys within RE enhances the software engineering process to have more sustainability and less negative influence on the environment.

While there are studies about using recommender systems in requirements engineering, only a few researches specifically focused on the Elicitation phase. For instance, Portugal et al. (Portugal et al., 2017) used GitHub projects as their source of information for implementing their recommender system to help categorize similar projects based on their domain concepts. Castro-Herrera et al. (Castro-Herrera et al., 2009) considered the challenges of stakeholders' collaboration in large-scale software projects; to overcome the issue, they introduced a semi-automated requirements elicitation framework that employs data-mining methods and a hybrid recommender system. AlZu'bi et al. (AlZu'bi et al., 2018) considered recording the current elicited requirements and, based on the Apriori algorithm, recommend the relevant requirements in similar future projects. Palomares et al. (Palomares et al., 2018), by presenting the OpenReq approach and building a recommender system, tried to help users in making better decisions and also enhance the quality assurance processes. There are also two noticeable highly relevant researches to our work entitled StakeNet (Lim et al., 2010) and StakeRare (Lim & Finkelstein, 2011). StakeNet provided a social network and asked users to recommend other stakeholders to prioritizes them by considering their role and importance on the project (Lim et al., 2010). StakeRare, an improved system based on the StakeNet, can be considered as the most similar study to our work; the authors introduced their method as a system for requirements elicitation in large software projects. In StakeRare, after building a social network for prioritizing stakeholders, the system asked them to rate some predefined requirements. StakeRare finally considers requirements' scores and runs a collaborative filtering-based recommender system to suggest related requirements (Lim & Finkelstein, 2011).

Although mentioned studies like the StakeRare have a similar approach with our work in using the benefits of recommender systems in the requirement elicitation activities, the Req-Rec has some fundamental and valuable differences and contributions:

- While our system functionality is tested to elicit requirements of a university enrollment system, it can also consider various types of requirements from security to any kind of non-functional requirements. There is no limitation about the project domain range.
- The repertory grid can be ended entirely online for 24 hours without any time limitation, likewise, remotely without any location-based restrictions. Moreover, the system can be automatically translated into different languages for international projects.
- Unlike content filtering-based methods, there is no dependency on the data type. The Req-Rec only considered stakeholders' IDs, their ratings and elicited requirements as items with IDs.

The following paragraphs include requirements elicitation techniques and approaches presented in Fig. 2. The evaluation of the mentioned techniques is summed up in Table 1. Finally, at the end of this section, reasons are explained for why the repertory grid technique is chosen as the fundamental element of the Req-Rec.

### 2.1. Requirement elicitation techniques and approaches

As mentioned before, requirement elicitation is a critical phase in requirement engineering. There are many different methods for this phase's tasks about identifying stakeholders and interacting with them. Each elicitation technique is explained separately by considering its strengths and weaknesses in the coming sections.

#### 2.1.1. Traditional techniques

In comparing all the four main types of introduced elicitation techniques in Fig. 2., Traditional techniques are the most basic and commonly used ones. Most types of these techniques are based on one-on-one interviews with stakeholders. Studied traditional techniques in this section are: interviews, document analysis, questionnaires and surveys, and introspection (Bradley et al., 2006; Yousuf & Asger, 2015).

- **Interviews**

  Personal interview (Agarwal & Tanniru, 1990; Holtzblatt & Beyer, 1995) with stakeholders probably is the most traditional method and is widely used as the primary technique to elicit requirements. The process of an interview is simple and generally includes four steps (Van Lamsweerde, 2009): Selecting stakeholders, Running a meeting, Writing a report from the results, Evaluating the achievements with stakeholders.

  The main weak point of the interview technique is that Interviewers need to have good communication skills and be able to engage in face-to-face discussions (Dick et al., 2017 ). Although there have been some achievements in human-computer interactions (HCI) for building automated systems to conduct interviews in recent years, arranging sessions with a group of people is still time-consuming. Therefore, in most of the presented systems, there is a lack of a believable impressive artificial character that is not just a digital face embodied by a human voice (Pickard & Roster, 2020). However, the interview technique has two significant and remarkable advantages, and they are easy to implement and cost-effective. Interviews can be categorized into three different types (Yousuf & Asger, 2015; Zowghi & Coulin, 2005):

  - **Structured**

    Pre-established with questions to asked from stakeholders.
  - **Semi-structured**

    The combination of a not wholly organized and unstructured interview.
  - **Unstructured**

    Unstructured interviews are the opposite side of structured ones. There are no pre-defined questions in these types of meetings (Wilson, 2014).

- **Document Analysis**

  This technique is also known as background study or content analysis. It is highly related to human evaluation and confirmation, so it is inconsistent with the time factor. The main disadvantage of document analysis is data mining issues on huge amounts of data. Studying related information and analyzing materials are the two main activities in document analysis.(C. J. Davis et al., 2006; Van Lamsweerde, 2009).

- **Questionnaire and Surveys**

  Questionnaires (Foddy & Foddy, 1994) and surveys are cost-effective ways to gather information through a series of codified questions. They provide the possibility to reach valuable individuals' subjective opinions and ideas, which are not easily accessible from task performance elicitation techniques alone (Clark & Maguire, 2020; Wellsandt et al., 2014). The negative side is that they can be time-consuming to prepare high-quality questionnaires. Questions can be asked in multiple-choice or weighting question style and asked through papers or online platforms

- **Introspection**

  In the introspection technique (Goguen & Linde, 1993), analysts elicit the requirements based on their ideas about stakeholders' needs. Due to individuals' knowledge limitations, the Introspection technique is usually used only in the first steps of the requirements elicitation phase. The method can be helpful when the requirement engineer is an expert on the intended field.

### 2.1.2. Contextual techniques

Contextual techniques refer to observational requirements elicitation methods that are done at the customer’s workplace. Observation and ethnography are the most well-known types of contextual techniques (Sharma & Pandey, 2014).

- **Observation**

  Observation means doing researches about stakeholders' requirements in their natural conditions. This technique is also called social analysis. Observation is often done with other elicitation techniques, such as interviews (Yousuf & Asger, 2015). The main interest of the observation technique is its capacity for discovering tacit knowledge and hidden concerns that can not easily understand with other methods. While observation can sometimes be considered an inexpensive method, it is highly related to the time limitation issue. Another challenge in gathering information by observation is users often act and do their tasks differently when there is an observer in their environment (Younas et al., 2017).

- **Ethnography**

  In requirement engineering, ethnography refers to the study of the stakeholders by observing the target society in their natural setting for a while. It provides the opportunity to experience some unexpected issues, and this feature helps the observer to comprehend the stakeholders' culture. However, getting an impressive result from ethnography can be time-consuming, and there is a high probability of biased results (Zowghi & Coulin, 2005).

### 2.1.3. Collaborative/Groups techniques

In collaborative or group techniques elicitation methods, requirements are collected by more than one person in face-to-face communication. Some of the critical methods of this category are prototyping, joint application development (JAD), brainstorming, requirements workshops, focus groups (group work), and user scenarios. As a consequence of recent online improvements, some of the collaborative techniques such as brainstorming and negotiations can be done online (Iqbal et al., 2019; Qi et al., 2016).

- **Prototyping**

  Prototype (Davis, 1992) such as dummy versions, pictures, or sketches are used in the early development life cycle steps to make stakeholders more familiar with considered solutions or products. Prototypes are considered throwaway prototypes if used only once and evolutionary if built to develop into the final product (Younas et al., 2017).

- **Joint application development (JAD)**

  JAD refers to structured meetings where all related parties come together. As a result of the knowledge combination, participants can explore a variety of requirements, then review them and finally rate them (Jackson & Embley, 1996). The significant difference between JAD sessions and brainstorming meetings are roles for cooperators, the central purposes of the session, and establishing all the actions before stakeholders participate in the session.(Browne et al., 2018).

- **Brainstorming**

Brainstorming is such a useful technique, specifically at the beginning of a novel project. As all the participants were encouraged to share their opinions without any concern of judgment. Generally, brainstorming sessions are like unstructured meetings in the observation technique, which has been discussed before (Bonnardel & Didier, 2020).

- **Requirements workshops**

  Requirements Workshops are structured group meetings and are somehow similar to JAD (Yousuf & Asger, 2015; Zowghi & Coulin, 2005). They have defined roles, such as facilitator, content participant, recorder, observer, on-call subject matter expert, and workshop sponsor. The facilitator is responsible for planning and leading the process and also suggests requirements deliverables (Gottesdiener, 2003).

- **Focus groups (group work)**

  Focus Groups are one of the swift methods used for eliciting and purifying requirements for the project. To increase the performance of the session, usually, there is a limitation for the number of participants, and each meeting typically includes around 6 to 12 members (Ramdhani et al., 2018).

- **Focus groups (user scenarios)**

  Scenarios show the way that systems interact with users through stories or samples from the real world (Moallemi et al., 2019). They are so beneficial for requirements evaluation and test cases. The first activities about using Scenarios as a technique for requirement elicitation in RE came back to the 1980s (Jarke et al., 1998; Rafiq et al., 2017). Scenarios based on their concept can be categorized into the two following types (Van Lamsweerde, 2009):

  - **Positive Scenarios**

    A Positive scenario explains what the system should do when the stakeholder ends a request about something.

  - **Negative Scenarios**

    A negative scenario indicates what may not occur when a specific situation happens, and the system has to exclude some predefined rules.

### 2.1.4. Cognitive techniques

Most cognitive techniques such as conceptual laddering, card sorting, repertory grid, and class responsibility collaboration (CRC) can be considered as artifact driven requirement elicitation methods (Van Lamsweerde, 2009). The requirement elicitation methods, which need some tools to extract the knowledge and can be easily done by a face-to-face interview or a simple observation (Rietz & Maedche, 2019).

- **Conceptual laddering**

  The main idea behind the laddering technique is using the concept of the taxonomical tree to elicit stakeholders' requirements. To draw the hierarchical tree, stakeholders are asked a set of questions to classify related requirements (Hinkle, 1965). The laddering technique can be done by specific tools or by a specialist. The laddering technique on the plus side is simple, cheap, and easy to implement. However, if the target systems have too large requirements, it will be complex to define relevant information and draw the taxonomy. Furthermore, modifications are another issue in complex trees because one change can make a series of other changes. Also, there is a risk of too subjective and not accurate results (Elijah et al., 2017).

- **Card sorting**

  Like the laddering technique, the primary use of card sorting is to elicit requirements and classifying them (Morente-Molinera et al., 2019). In this technique, stakeholders are provided with some cards that each of them presents a domain entities' names textually or graphically. Stakeholders then asked to arrange and group similar cards based on their ideas and understanding (Rao & Katz, 1971).

- **Repertory grid**

  The repertory grid was first introduced by George A.Kelly in 1991 (George, 1991). The main component of this technique is a matrix. Rows of the matrix known as Constructs and columns indicate Elements. Each row has two Constructs on the right side and left side. Constructs of each of the rows have the opposite meaning of each other. Stakeholders are asked to rate each element by considering the Constructs. Ratings near 0 mean that the stakeholder is more satisfied with the left Construct, while scores near 5 are closer to the right Constructs (Dey & Lee, 2017). Fig. 3. shows a blank repertory grid. Compared to other cognitive requirement elicitation techniques like card sorting, the repertory grid provides more precise results. However, a repertory grid can not be efficient for complex systems. Moreover, filling the cells and generally completing the matrix may be time-consuming.

| | **Element 1** | **Element 2** | **Element 3** | |
|---|---|---|---|---|
| Construct 1 | | | | Construct 1 |
| Construct 2 | | | | Construct 2 |
| Construct 3 | | | | Construct 3 |
| ⋮ | | | | ⋮ |
| Construct n | | | | Construct n |

**Fig. 3. The structure of the repertory grid.**

- **Class responsibility collaboration (CRC)**

The class responsibility collaboration (CRC) technique is one of the oldest methods for requirement elicitation. CRC is a derivative of card sorting introduced by Kent Beck and Ward Cunningham in 1989 (Beck & Cunningham, 1989). The technique focuses highly on object-oriented systems and is a valuable mechanism for conceptual modeling and detailed design (Keller et al., 2019). Fig. 4. shows a sample of a CRC card. The main elements of CRC are cards that generally express software requirements. Each card refers to a defined class and is arranged based on its functionality and collaborations with other classes. Requirements are then formulated using these cards (Yousuf & Asger, 2015).

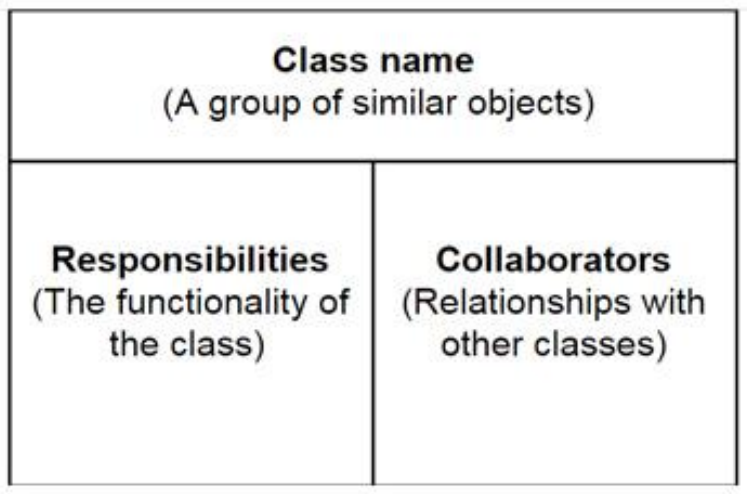

**Fig. 4. A sample of a CRC card.**

- **Protocol analysis**

T In this technique, stakeholders are asked to think loudly and describe their current actions while an observer tries to understand the situation and elicit the related requirements (Ericsson & Simon, 1998). Based on this concept, the protocol analysis technique also can be a kind of passive observation (Heirweg et al., 2019).

**Table 1. An overview of the most well-known requirement elicitation techniques and their strengths and weaknesses.**

| Category | Technique or Approach | Strength | Weakness |
|---|---|---|---|
| Traditional | Interview | ▪ Easy to implement<br>▪ Cost-effective | ▪ Needs to be done face-to-face<br>▪ Needs psychological skills such as effective communication<br>▪ Time-consuming |
| | Document Analysis | ▪ Gathering useful information even for other phases of the project<br>▪ Requirement engineer becomes more familiar with the system and its stakeholders | ▪ A large amount of data that have to be considered.<br>▪ The magnitude of unrelated and irrelevant information<br>▪ Outdated and inaccurate resources<br>▪ Needs lots of time |
| | Questioner / Survey | • Quick<br>• Cheap<br>• Easy for stakeholders | • Time-consuming to high-quality questionnaires<br>• Lack of participants<br>• Chance of ambiguity for answering<br>• Bias questions |
| | Introspection | • Easy to execute<br>• Cheap<br>• Quick | • Limitation of knowledge<br>• Incompleteness |
| Contextual | Observation | • Revealing tacit knowledge<br>• Exploring hidden problems | • Highly related to the time limitation issue.<br>• Users act differently in observation sessions |
| | Ethnography | • Covering unexpected issues<br>• Understanding stakeholder culture | • Time-consuming<br>• Too subjective |
| Collaborative/Groups | Prototyping | • Helps stakeholders to get a better understanding of the product<br>• Finding hidden needs<br>• Saving time and financial resources in the development and implementation phases | • Time-consuming itself<br>• Expensive<br>• Increase expectations too much |
| | Joint Application Development (JAD) | • Exploring variety of requirements<br>• Low price | • Potential to be Time-consuming<br>• Not practical for significant issues |
| | Brainstorming | • Increasing creativity<br>• Low price<br>• Easy to implement | • Challenge of quality vs. quantity<br>• People have different personalities<br>• Not practical for significant issues |
| | Requirements Workshop | • Saving time<br>• Low price | • Difficult to organize because of conflicts in stakeholder's schedule<br>• Lots of participants may slow down the process |
| | Focus Group (Group Work) | • Saving time<br>• Low price<br>• Impressive Results | • Difficult to organize because of conflicts in stakeholder's schedule<br>• User act differently in observation sessions |
| | Focus Group (User Scenario) | • Clarifying standard flow, uncommon conditions, and alternative solutions | • Risk of over-specification<br>• Including irrelevant details<br>• High probability incompleteness |
| Cognitive | Conceptual Laddering | • Low price<br>• Easy to use<br>• Easy to implement | • Not efficient for complex systems<br>• Subjective<br>• Inaccurate<br>• Irrelevant results |
| | Card Sorting | • Low price<br>• Fast<br>• Easy to use<br>• Revealing tacit knowledge<br>• In-depth recognition of the stakeholder's thinking mode | • Not efficient for complex systems<br>• Subjective<br>• Inaccurate<br>• Irrelevant results |
| | Repertory Grid | • Acceptable precision<br>• Balancing the differences between Stakeholders' and experts' points of view | • Not efficient for complex systems<br>• Time-consuming |
| | Class Responsibility Collaboration (CRC) | • Low price<br>• Easy to use<br>• Cause collaboration between experts and stakeholders<br>• A proper complement for UML diagrams | • Not efficient for large and complex systems<br>• Working with too many cards may be time-consuming |
| | Protocol Analysis | • Revealing tacit knowledge<br>• Low price<br>• Easy to implement | • Time-consuming<br>• Results may be incomplete and inaccurate |

By considering all of the mentioned techniques and analyzing their pros and cons, it can be concluded that the repertory grid method is a reasonable way to clearly discover what stakeholders want while evading perceptual bias in the requirements' elicitation process (Edwards et al., 2009; Shaw, 1980). Many studies used the repertory grid as an efficient technique to understand stakeholders' needs in different fields, from inpatient care (Wittkowski et al., 2019) to setting information security policy in e-commerce (Samonas et al., 2020). The most considerable advantage of using the repertory grid requirements' elicitation process compared to other mentioned techniques can be summarized as acceptable precision in results and balancing the differences between stakeholders' and experts' points of view. Moreover, enhancing the repertory grid by adding dynamic, highly related recommendations can significantly overcome the time limitation and location-based restrictions and greatly assist stakeholders in becoming more aware of different aspects of the project.

## 3. Req-Rec Method

Regarding what has been mentioned above about the repertory grid advantages and the high similarity between this elicitation technique with the user-item rating, as the fundamental element of the collaborative filtering recommender systems, the Req-Rec method is proposed. The architecture is shown in Fig. 5. In the requirement engineering four steps concept (Van Lamsweerde, 2009), the result of the grid after doing analysis will be the input of the second step, which is about negotiation for the extracted requirements and trying to find solutions for them.

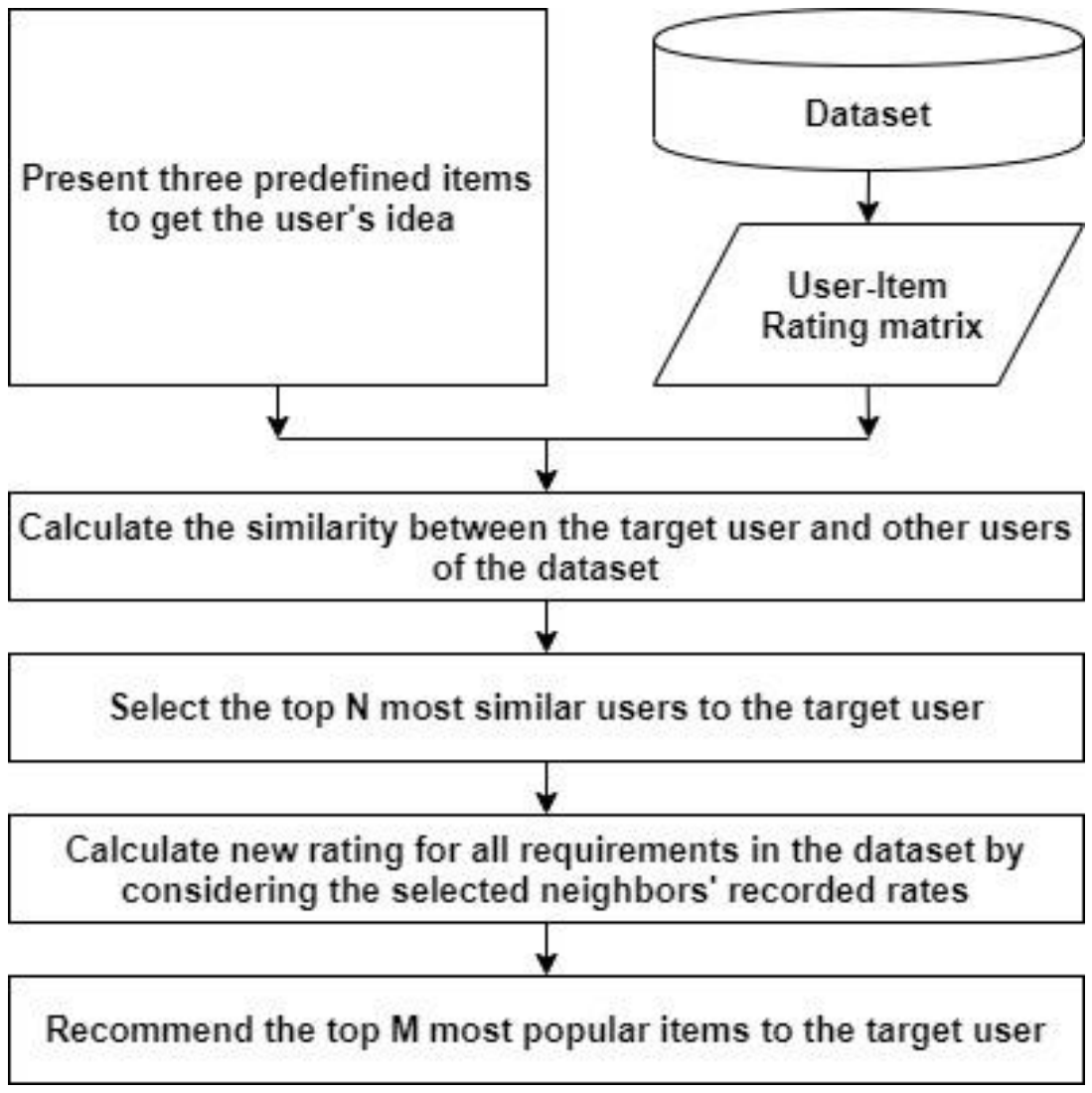

**Fig. 4. An overall view of the Req-Rec.**

The Req-Rec is clarified step by step in Algorithm 1.

**Algorithm 1 Req-Rec**

```
Input Requirements' recommendation (N, M, K)
Output Final recommendations
Begin
Presenting N requirements from the dataset to the target user
for i in range(0, length(number of users of the dataset)) do
        //Calculating the similarity between the target users and other users in the dataset by using the Pearson formula
    SimilarityAarray ← PCC(target user ,i^th user of the dataset)
SimilarityAarray = Sorted(SimilarityAarray)
for j in range(0, M) do
        //Selecting top M users of the SimilarityAarray
    TopUsers ← SimilarityAarray M^th user
for l in range(0, length(number of items of the dataset)) do
        //Predicting the user's rating score for all the requirements in the dataset by using the Resnick formula
    PredictedRates ← Resnick (target user, l^th item,  TopUsers)
for t in range(0, PredictedRates) do
        //Selecting top K items with the highest predicted rate from the PredictedRates
RecommendationCandidates ← PredictedRatest^th item
Recommending RecommendationCandidates to the target user
End Algorithm
```

1. First, the system presents ***N*** predefined requirements to the user to understand the user's concerns.
2. Second, proving the user-item rating matrix from the dataset.
3. Next, calculate the similarity between the target users and other users in the dataset. To aim this goal, the Pearson Correlation Coefficient (Pindyck et al., 1991) formula is used as the current most popular similarity measure for building a collaborative filtering recommender system (Leskovec et al., 2020).

   The Pearson Correlation Coefficient (PCC) formula considers the corated items' ratings to measure the similarity between the two target users. Formula (1) shows the PCC formula.

$$PCC_{a,b} = \frac{\sum_{i=1}^{I_{a,b}} \left(r_{a,i} - \bar{r}_a\right)\left(r_{b,i} - \bar{r}_b\right)}{\sqrt{\sum_{i=1}^{I_a} \left(r_{a,i} - \bar{r}_a\right)^2} \sqrt{\sum_{i=1}^{I_b} \left(r_{b,i} - \bar{r}_b\right)^2}} \qquad (1)$$

In PCC formula, $r_{a,i}$ means rating score for item $i$, from the target user a, and $r_{b,i}$ respectively means the rating score to item $i$ from the user $b$. $\bar{r}_a$ and $\bar{r}_b$ are also show the average rating of user $a$, and user $b$ based on all rated items by each user.

4. Then, the top M most similar users to the target user are selected.
5. After that, the proposed system predicts the possible target user's rating score for all the requirements in the dataset. To make the predictions, Resnick's standard prediction (Resnick et al., 1994) is used. Formula (2) shows the Resnick formula.

$$a(i) = \frac{\sum_{b \in B(i)} \left( \mathrm{b}(\mathrm{i}) - \bar{b} \right) PCC_{a,b}}{\sum_{b \in B(i)} \left| PCC_{a,b} \right|} \qquad (2)$$

In the above formula, $a_{(i)}$ is the rating to be predicted for item $i$ by the target user $a$. $b_{(i)}$ is the rating for item $i$ by the user $b$. $\bar{r}_a$ and $\bar{r}_b$ mean the average of rating scores for users $a$ and $b$. $PCC_{a,b}$ is the similarity between the users $a$ and $b$ and output of the Formula (1).

6. Finally, the algorithm provides its recommendation by the top $K$ requirements with the highest rating score to the target user.

## 4. Experimental Results

The Req-Rec method was conducted and evaluated in two printed and online versions. The printed version was the first activity and done for 10 Master's degree students in computer software engineering, who passed the Requirement Engineering course. The proposed repertory grid is shown in Fig. 6.

NO:

Subject: Requirement elicitation for a university enrollment system by using a collaborative filtering recommender system

**Hi,**

**Thanks for your collaboration in doing this research. You can find an explanation about feeling this table at the bottom of the page.**

| Construct | Rate ← → | | | | | | | | | | | Construct |
|---|---|---|---|---|---|---|---|---|---|---|---|---|
| | | | | | | | | | | | | |
| | | | | | | | | | | | | |
| | | | | | | | | | | | | |
| | | | | | | | | | | | | |
| | | | | | | | | | | | | |

Example: the provided table has the structure of the Repertory Grid. The subject of the research is about eliciting the requirements of a university enrollment system. The following table is an example of some of the most common Construct of the subject.

| Construct | Rate ← → | | | | | | | | | | | Construct |
|---|---|---|---|---|---|---|---|---|---|---|---|---|
| Low level of security | | * | | | | | | | | | | High level of security |
| Poor user interface | | | | | | * | | | | | | Excellent user interface |
| Low speed | | | | | | | | | | | * | High speed |

It can be concluded from the ratings that High speed is the most important factor for the target user. Security was not a priority, and the User interface has the least importance.

**Fig. 6. The proposed repertory grid.**

Then to improve the proposed system's evaluation, an online version is implemented. During about the two months, as a result of collaboration from professors, university students, and some expert computer scientists and software engineers, the number of users increased from 10 to 50. Then a module was added for recording user's feedback about the recommendations. The feedback gathering module also was online for two months. Totally 127 participants, including different educational levels, were associated in this phase.

### 4.1. Data Gathering

After achieving the 50 participants from the paper-based and online form, irrelevant data is omitted and similar ones are also combined. As a result of these activities, all of the defined constructs were organized into the 12 independent constructs, which were:

- Reliability of the system. (less technical errors and more up-time of the server)
- Professor's information (their performance in the last semesters.)
- Ability to reserve courses when the classes are full.
- Online support and enough "How to use" content, especially for new features and changes.
- Effective filtering, based on different factors such as days, locations, contents, etc.

- Privacy
- User-friendliness (using User Experience (UX) rules, such as proper color for the buttons with critical functionality.)
- Speed and performance
- Responsive layout in different screen sizes like mobiles and tablets.
- Accurate and precise online data
- Easy to use
- Cross-platform (work independently from operating systems and browsers.)

### 4.2. System evaluation

To evaluate the proposed method and users' satisfaction with the provided recommendations, a module is added to record the stakeholder's feedback about the recommendations in the online version of the system. The module was based on the star rating system, which 0 means no idea, and 5 stars imply practical and perfect recommendations. By considering what has been mentioned in section 3 about the architecture of the Req-Rec, the main variables in this section are:

- N = 3
- M = 5
- K = 5

Respectively, N represents the number of items shown to the users at the beginning step. M indicates the number of the most similar users to the target user. K shows the number of requirements that have the highest rating score to the target user. Fig. 7 shows an overview of how the proposed system components work, and its functionality is described as follows:

1. First, the system presents 3 predefined requirements for the user to understand the user's concerns and priorities. Based on the data-gathering phase, the 3 most popular requirements (rated by most users) about the university enrolment system were:
   1. reliability of the system, 2. professor's information, 3. ability to reserve courses. In this scenario $N = 3$. Educational level (Ph.D., Master, Bachelor) also was asked in this step.
2. Second, the recorded data is used to prepare the inputs of the collaborative filtering module in the next step.

3. Next, the Pearson Correlation Coefficient is used to calculate the similarity between the target users and the other core 50 users in the dataset. The module itself has two connections with the Average rating calculator and the Finding corated requirements modules.
4. Then, the top ***M*** most similar users to the target user are selected. In this step ***M*** = 5.
5. After that, by using Resnick's standard prediction, the system predicted the possible target user's rating score for all other 9 the requirements in the dataset.
6. Finally, the system recommended the top ***K*** requirements with the highest rating score to the target user. In this section ***K*** = 5.

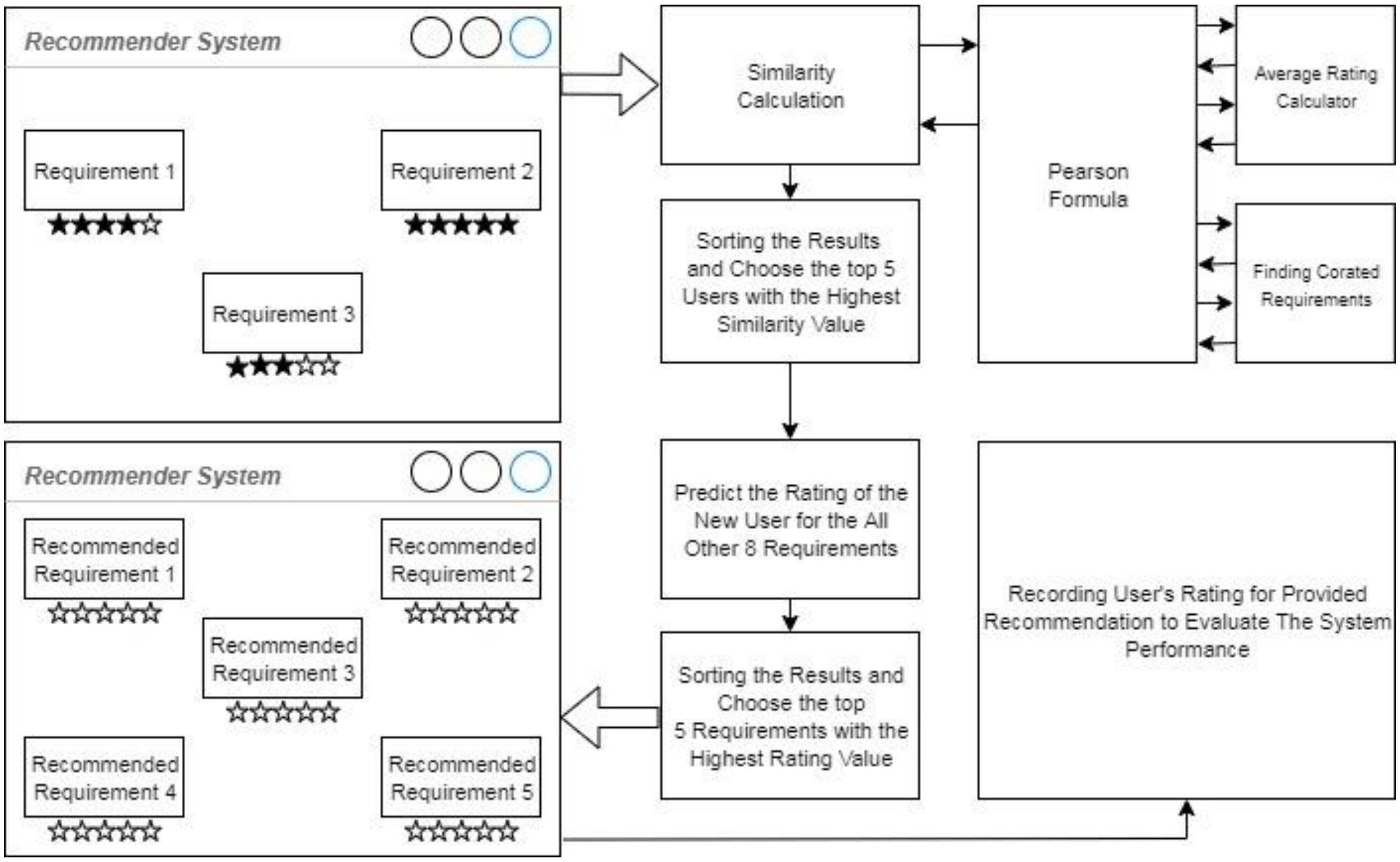

**Fig. 7. The view of the proposed recommender system functionality.**

As mentioned before, 127 users were asked to show their satisfaction with the recommended requirements in the scenario of elicitation requirements of a university enrollment system in a star rating system, in which 0 were no idea and rates from 5 to 1 were their satisfaction in descending. Clearly, the Req-Rec could satisfy its stakeholders with an average of more than 50 percent (about 57) by recommending relevance requirements. Details of the participant in the evaluation phase, shown in table 2.

**Table 2. Details of the participant in the evaluation phase.**

| Education level | **Ph.D.** | **Master** | **Bachelor** |
|---|---|---|---|
| Number of participants | 60 | 46 | 21 |
| Total number of participants | | 127 | |

The chart in Fig. 8 indicates the average of users' feedbacks about the elicited requirements. The horizontal axis shows the participants' educational levels, including Ph.D., Master, Bachelor, and the vertical axis is the average score based on the stakeholders' feedback for the final five recommended requirements.

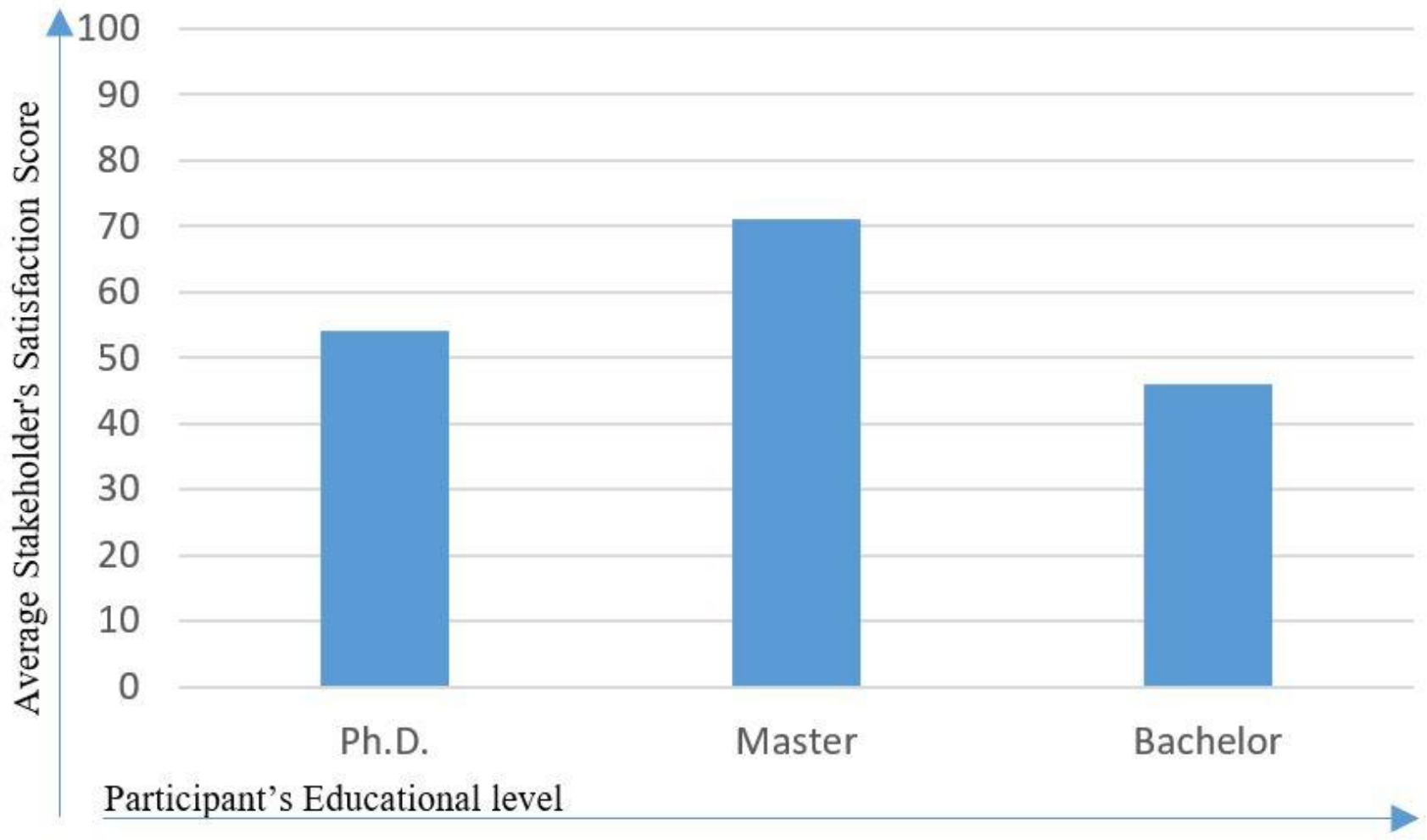

**Fig. 7. The result of the stakeholder's satisfaction feedbacks analysis.**

## 5. Conclusion and Future Works

This work's primary goal was to propose a suitable solution for increasing stakeholder's satisfaction by helping them in requirements elicitation. The repertory grid technique was used as the fundamental element for data gathering and the most critical requirement elicitation. Then, based on the collaborative filtering approach, a recommender system was built to efficiently assist stakeholders in exploring different aspects of the project and overcoming the time-consuming challenge. The results admit the effectiveness of the Req-Rec to improve stakeholder's satisfaction.

The main limitation of this research was the lack of the number of available participants. However, even though additional tests should be performed on a larger dataset, it seems realistic that the Req-Rec method could be used successfully as an efficient requirement elicitation technique.

In future work, unsupervised clustering techniques will be used for categorizing functional and non-functional requirements and recommend each type of requirement separately. These kinds of segmentation can help identify relevant groups of stakeholders and prioritize related requirements.

### Competing interests

We allude that there exists no dispute of interest between authors.

### Authors' contributions

All three authors contributed to the research work and to writing the manuscript.